# A Secure Experimentation Sandbox for the design and execution of trusted and secure analytics in the aviation domain

Dimitrios Miltiadou[1][0000-0001-9030-3299], Stamatis Pitsios[1][0000-0001-9525-1416], Dimitrios Spyropoulos[1][0000-0002-5820-4319], Dimitrios Alexandrou[1][0000-0002-2712-2089], Fenareti Lampathaki[2][0000-0002-3131-4622], Domenico Messina[3][0000-0001-7455-2623], Konstantinos Perakis[1][0000-0002-2236-9816]

[1] UBITECH, Thessalias 8 & Etolias, Chalandri, 15231, Greece
[2] SUITE5, Alexandreias 2, Bridge Tower, Limassol, 3013, Cyprus
[3] ENGINEERING Ingegneria Informatica S.p.A., Piazzale dell Agricoltura 24, Rome, 00144, Italy

**Abstract.** The undergoing digital transformation of the aviation industry is driven by the rise of cyber-physical systems and sensors and their massive deployment in airplanes, the proliferation of autonomous drones and next-level interfaces in the airports, connected aircrafts-airports-aviation ecosystems and is acknowledged as one of the most significant step-function changes in the aviation history. The aviation industry as well as the industries that benefit and are highly dependent or linked to it (e.g. tourism, health, security, transport, public administration) are ripe for innovation in the form of Big Data analytics. Leveraging Big Data requires the effective and efficient analysis of huge amounts of unstructured data that are harnessed and processed towards revealing trends, unseen patterns, hidden correlations, and new information, and towards immediately extracting knowledgeable information that can enable prediction and decision making. Conceptually, the big data lifecycle can be divided into three main phases: i) the data acquisition, ii) the data storage and iii) the data analytics. For each phase, the number of available big data technologies and tools that exploit these technologies is constantly growing, while at the same time the existing tools are rapidly evolving and empowered with new features. However, the Big Data era comes with new challenges and one of the crucial challenges faced nowadays is how to effectively handle information security while managing massive and rapidly evolving data from heterogeneous data sources. While multiple technologies and techniques have emerged, there is a need to find a balance between multiple security requirements, privacy obligations, system performance and rapid dynamic analysis on diverse large data sets. The current paper aims to introduce the ICARUS Secure Experimentation Sandbox of the ICARUS platform. The ICARUS platform aims to provide a big data-enabled platform that aspires to become an "one-stop shop" for aviation data and intelligence marketplace that provides a trusted and secure "sandboxed" analytics workspace, allowing the exploration, curation, integration and deep analysis of original, synthesized and derivative data characterized by different velocity, variety and volume in a trusted and fair manner. Towards this end, a Secure

Experimentation Sandbox has been designed and integrated in the holistic ICARUS platform offering, that enables the provisioning of a sophisticated environment that can completely guarantee the safety and confidentiality of data, allowing to any interested party to utilise the platform to conduct analytical experiments in closed-lab conditions.

## 1  Introduction

The Aviation industry encapsulates the design, development, production, operation and management of aircrafts with a wide range of products ranging from aircraft, rotorcraft, engines, avionics and systems to leading operations and services. The undergoing digital transformation of the aviation industry is driven by the rise of cyber-physical systems and sensors and their massive deployment in airplanes, the proliferation of autonomous drones and next-level interfaces in the airports, connected aircrafts-airports-aviation ecosystems and is acknowledged as one of the most significant step-function changes in the aviation history. It is estimated that an average flight can produce between 500 and 1,000 gigabytes of data [1], and according to estimations the global fleet could generate up to 98,000,000 terabytes of data by 2026 [2]. The aviation industry as well as the industries that benefit and are highly dependent or linked to it (e.g. tourism, health, security, transport, public administration), are ripe for innovation in the form of Big Data analytics.

Leveraging Big Data requires the effective and efficient analysis of huge amounts of unstructured data that are harnessed and processed towards revealing trends, unseen patterns, hidden correlations, and new information, and towards immediately extracting knowledgeable information that can enable prediction and decision making[3][4]. Big data technologies is a new generation of technologies that aims to add value to a massive volume of data with various formats by enabling high-velocity capture, discovery, and analysis [5]. Conceptually, the big data lifecycle can be divided into three main phases: i) the data acquisition, ii) the data storage and iii) the data analytics. For each phase, the number of available technologies and tools that exploit these technologies is constantly growing, while at the same time the existing tools are rapidly evolving and empowered with new features. However, the Big Data era, comes with new challenges and one of the crucial challenges faced nowadays is how to effectively handle information security while managing massive and rapidly evolving data from heterogeneous data sources. Inevitably, the data analytics and extraction of intelligence from them requires dynamic data sharing between different internal stakeholders of an organisation or even between external stakeholders, as well as data access to all these stakeholders. However, this introduces multiple security threats such as the discovery of private or confidential information and unauthorised access to data at storage or data in motion. While multiple technologies and techniques have emerged, there is a need to find a balance between multiple security requirements, privacy obligations, system performance and rapid dynamic analysis on diverse large data sets [6].

The current paper aims to introduce the ICARUS Secure Experimentation Sandbox of the ICARUS platform. The ICARUS platform aims to provide a big data-enabled platform that aspires to become an "one-stop shop" for aviation data and intelligence marketplace that provides a trusted and secure "sandboxed" analytics workspace, allowing the exploration, curation, integration and deep analysis of original, synthesized and derivative data characterized by different velocity, variety and volume in a trusted and fair manner. Towards this end, a Secure Experimentation Sandbox has been designed and integrated in the holistic ICARUS platform offering, that enables the provisioning of a sophisticated environment that can completely guarantee the safety and confidentiality of data, allowing to any interested party to utilise the platform to conduct analytical experiments in closed-lab conditions, in order to effectively and securely address the critical barriers for the adoption of Big Data in the aviation industry, and enable aviation-related big data scenarios for EU-based companies, organizations and scientists.

## 2     Materials and methods

In spite of the important developments in the big data technologies, an analysis of these technologies in respect to the adopted architectures and techniques revealed that many shortcomings in terms of security still exist [7]. Furthermore, the importance of security and privacy measures is increasing, along with the growth in the generation, access, and utilization of Big Data [8]. The typical characteristics of big data, namely velocity, volume and variety associated with large-scale cloud infrastructures and the Internet of Things (IoT) revealed the inadequacy of the traditional security and privacy mechanisms as they fail to cope with the rapid data explosion in such a complex distributed computing environment, as well as with the scalability, interoperability and adaptability of contemporary technologies that are required for big data [9]. At the same time, the number of malicious attacks against big data infrastructure is on the rise, as it was revealed by recent surveys focused on the security aspect of big data [10]. Gathering, storing, searching, sharing, transferring, analysing and presenting data as per requirements are the major challenging task in big data [11]. In any big data platform, the strategic priorities related to security should be clearly defined and the guidelines for choosing the associated technologies in terms of reliability, performance, maturity, scalability and overall cost should be also clearly established to ensure that the design platform provide the necessary security mechanisms that include, among others, the anonymisation of confidential or personal data, the data cryptography, the centralised security management, the data confidentiality and data access monitoring [6]. At the same time, it should ensure their future evolutions will be easily integrated in the existing solution.

Fortunately, the recent advancements and trends in the big data technologies and the adopted strategies provide a new compelling opportunity to design and build a big data platform that incorporates a holistic security approach capable of addressing the security and privacy challenges imposed by the nature of big data and the requirements of data analytics stakeholders.

## 2.1 The ICARUS technical solution

The main objective of the ICARUS platform [12] is to provide a multi-sided platform that that will allow exploration, curation, integration and deep analysis of original, synthesized and derivative data characterized by different velocity, variety and volume in a trusted and fair manner. Furthermore, the platform builds a novel data value chain in the aviation-related sectors towards data-driven innovation and collaboration across currently diversified and fragmented industry players, acting as multiplier of the "combined" data value that can be accrued, shared and traded, and rejuvenating the existing, increasingly non-linear models / processes in aviation.

The key objectives of the platform is to offer a scalable and flexible big data-enabled environment that provides secure and trusted: a) data preparation and data upload, b) data exploration, data sharing and data brokerage and c) data analysis execution and visualization generation capabilities. Security and privacy were considered as crucial pillars of the ICARUS platform. To this end, a security and privacy by-design approach was adopted in order to effectively and efficiently cover all the aspects related to the information protection and secure data management over the entire data lifecycle. In this context, the ICARUS platform incorporates advanced security mechanisms that offer methods for increasing the security, the privacy and the data protection across all tiers of the architecture, taking into account the aviation industry's needs, requirements and peculiarities with regards to the security of information. The key points of this approach is the adoption of the end-to-end encryption imposed in all the datasets that are stored in the platform, as well as a secure decryption process for the effective data sharing of datasets within the scope of the platform.

The ICARUS platform's architecture is a modular architecture that provides enhanced flexibility and is composed of a set of key components that are built on top of efficient and state-of-the-art big data infrastructure, technologies and tools, maximizing the benefits of their effective combination. In detail, the platform architecture is composed by 22 key components that have been designed with the aim of delivering specific business services with a clear context, scope and set of features (see Fig.1).

The components are conceptually organised in three main tiers, the **On Premise Environment**, the **Core Platform** and the **Secure and Private Space**. Each tier is undertaking a set of functionalities of the platform depending on the execution environment and context. The scope of the On Premise Environment is to provide the required services that will perform all the data preparation steps as instructed by the Core Platform. In this context, the On Premise Environment is composed by multiple components that are running on the data provider's environment with the main purpose to prepare the data provider's private or confidential datasets in order to be uploaded in the platform. The On Premise Environment undertakes the responsibility of performing the tasks according to the instructions provided by the Core Platform.

The scope of the Core Platform is to provide all the required components for the execution of the core operations of the platform, as well as compilation of the instructions that are executed by the On Premise Environment and the Secure and Private Space. The Core Platform is composed of multiple interconnected components

5running on the platform's cloud infrastructure. It performs all core operations of the platform while also orchestrating and providing the instructions that are executed by the On Premise Environment and the Secure and Private Space. Furthermore, the Core Platform provides the only user interface of the platform as the rest of the tiers are incorporating only backend services.

The scope of the Secure and Private Space is to provide all the required components for the formulation of the trusted and secure advanced analytics execution environment of the platform. In order to cope with the emerging security and privacy requirements, the Secure and Private Space is providing a trusted and secure sandboxed analytics workspace supporting the existing rich stack of analytics tools and features, while at the same time providing strong security guarantees towards data confidentiality and data privacy. In this context, the Secure and Private Space contains a set of interconnected components that constitute the advanced analytics execution environment of the platform, whose management and orchestration is performed through the Core Platform. It provides the trusted and secure environment where the data analysis executed in accordance with the analytics workflow that is designed by the user within the Core Platform. The designed workflow is translated into a set of instructions which are executed by the responsible deployed components.

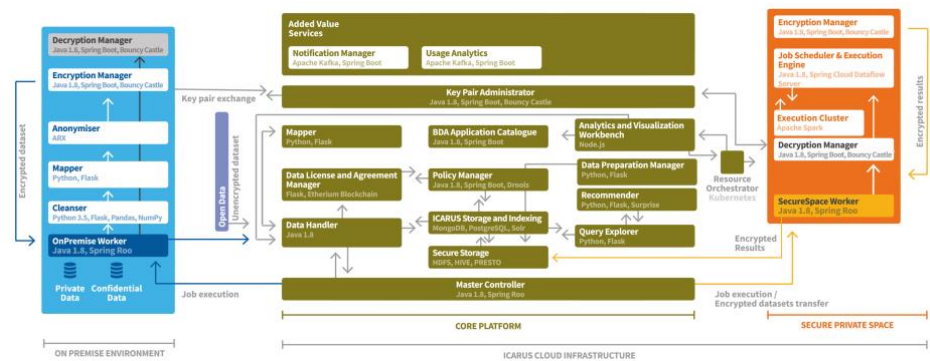

**Fig. 1.** The ICARUS platform conceptual architecture

The following section, focuses on the ICARUS Secure Experimentation Sandbox of the ICARUS platform's architecture that constitutes the trusted and secure sandboxed analytics workspace of the platform which facilitates the secure accessing and processing of big data towards the execution of advanced data analytics and visualisations over a modular and scalable architecture. The ICARUS Secure Experimentation Sandbox constitutes a cross-tier feature of the platform, that is orchestrated and controlled by the Core Platform tier and executed by the Secure and Private Space tier.

## 2.2   The ICARUS Secure Experimentation Sandbox

Data analytics rely on the effective aggregation and correlation of diverse multi-source data in order to generate new insights and knowledge. However, they are usually offered in conventional and in most cases through strictly defined fixed queries, constraining the fantasy of end-users and their experimentation potential which could lead to new knowledge insights in unexplored dimensions. At the same time, in the data storage and the data analytics phases of the big data lifecycle several security and privacy challenges arise as described in section 2.1 which is imperative that they should be properly addressed so that data analytics will be embraced by the organisations.

The ICARUS platform is by design supporting data security and privacy principles, to safeguard personal data, but also business critical data. Towards to this end, one of the core features of the platform is the provisioning of a sophisticated environment that can completely guarantee the safety and confidentiality of data, allowing to any interested party to utilise the platform to conduct analytical experiments in closed-lab conditions with the offering of the ICARUS Secure Experimentation Sandbox. The ICARUS Secure Experimentation Sandbox architecture holds a key role in the platform's solution, covering the needs for data analysis execution and visualization generation in a secure and trusted manner, effectively addressing the security and privacy challenges imposed by the nature of the big data, capitalising on the latest advancements in the techniques and technologies that enable easy, fast and secure deployment of containerised execution environments.

The design of the ICARUS Secure Experimentation Sandbox is composed by a set of components residing on the Core Platform tier, namely the Resource Orchestrator, the Secure Storage, the Analytics and Visualisation Workbench and the Data Preparation Manager, as well as all the components that formulate the Secure and Private Space tier, namely the SecureSpace Worker, the Job Scheduler and Execution Engine, the Encryption Manager and the Decryption Manager. In the following paragraphs, all the core components that are utilised in the realisation of the ICARUS Secure Experimentation Sandbox, as well as their interactions, are presented in detail.

**Resource Orchestrator**

The ICARUS Secure Experimentation Sandbox provides the deployment of a scalable isolated environment running on virtualised infrastructure and is based on technologies that enable easy, fast and secure deployment of containerised execution environments over virtualised infrastructure. The concept of containerised execution environments besides the portability and interoperability features, it also enables the monitoring, autoscaling and management of the deployed applications with the proper orchestration support. To this end, at the heart of the ICARUS Secure Experimentation Sandbox lays Docker, which is the container technology with the widest adoption by both the research and vendor communities. Docker containers are lightweight, standalone, self-contained systems that include everything that is needed for the proper execution of the system on a shared operating system such as code, runtime, system tools, system libraries and settings in an isolated manner. The container orchestration is handled by Kubernetes, which is also the most dominant open source platform for container orchestration, offering advanced monitoring capabilities, autoscaling and state-of-the-art management of deployed applications. Kubernetes encapsulates the concept of container virtualization thus enabling the deployment of containerized applications over virtualised infrastructure with extended orchestration support. Kubernetes is offering an additional layer of abstraction on top of the virtualised infrastructure providing out of the box service discovery and load balancing, storage orchestration, automated rollouts and rollbacks, automatic bin packing based on the available resources, health monitoring and self-healing mechanisms, as well as security and isolation between the deployed applications. The **Resource Orchestrator** is the component responsible for the provisioning of the Secure and Private Space, performing the deployment and monitoring of the required containerised services by exploiting the capabilities of Kubernetes and Docker.

**Secure Storage**

The **Secure Storage** undertakes the role of the private and secure storage of the ICARUS Secure Experimentation Sandbox. In this context, the Secure Storage component has a two-fold purpose: a) to store the data provider's private or confidential datasets, as well as the datasets acquired via the marketplace of the platform, prior to being used during the data analysis execution by the Secure and Private Space services and b) to store the data generated as a result of a data analysis performed by the Secure and Private Space services. In this process, the Secure Storage provides the private storage "spaces" with restricted access only to the owner of the "space". The design of the Secure Storage aims to better address the user requirements for the effective and efficient storage of datasets with high availability and high performance for the computationally-intense data analysis operations. Hence, the Secure Storage is composed by the Hadoop Distributed File System (HDFS), provisioned and monitored by the Apache Ambari, and the Apache Hive supported by the Presto query engine. The HDFS is utilised for the storage of the datasets that will be utilised in the data analysis. Apache Hive is operating on top of the HDFS as a data warehouse software that enables an abstraction layer on top of the datasets residing in HDFS, facilitating the complete data lifecycle management of the datasets, as well as providing an SQL-like interface enabling data query and analysis functionalities. Presto is offering the required high-performance query engine on top of the datasets by performing parallel query execution over a pure memory-based architecture running on top of Hive, enabling fast analytics queries against data of any size that are residing in Hive and consequently HDFS.

The SecureSpace worker residing in the Secure and Private Space is interacting with the Core Platform via the Master Controller to fetch the required dataset, that are afterwards decrypted by the Decryption Manager and are finally stored in the Secure Storage. Thus, the Secure Storage is providing the means to the Analytics and Visualisation Workbench component in order to access the transferred datasets and utilise them in the data analysis in an efficient and effective manner. Furthermore, the data generated in the Secure and Private Space upon the completion of a data analysis are encrypted by the Encryption Manager and provided through the SecureSpace Worker to the Secure Storage for storage. The encrypted results are fetched by the SecureSpace Worker and decrypted by the Decryption Manager in order to be provided as input for the visualisation process that is executed by the Analytics and Visualisation Workbench. Finally, the Secure Storage is enabling the data preparation operations performed by the Data Preparation Manager by enabling the access and manipulation of multiple datasets residing on the Secure Storage within the owner's private storage "space".

**Analytics and Visualisation Workbench**

The **Analytics and Visualisation Workbench** is the component enabling the design, execution and monitoring of the data analytics workflows within the platform and also where the visualisation and dashboards are displayed. The users are able to select an algorithm from the extended list of supported algorithms and set the corresponding parameters according to their needs. Furthermore, the users are provided with scheduling capabilities for the execution of the designed analytics workflow in a selected date and time. In terms of design, the Analytics and Visualisation Workbench consists of three main sub-components: (a) a very intuitive graphical user interface that allows users to exploit a set of machine learning algorithms to obtain meaningful information about his/her own data and display them with a set of built-in charts; (b) the repository for the implemented algorithms and (c) a microservice that exposes a RESTful API that enables the mediation between the frontend and the Job Scheduler and Execution Engine and the Resource Orchestrator backend components that the Analytics and Visualisation Workbench interacts with. As the nature of the component includes both a graphical user interface and a set of backend functionalities, different technologies are utilised for each purpose adopting the backend-for-frontend design pattern. The graphical user interface is mainly based on TypeScript, JavaScript, CSS and HTML5, while for the backend functionalities of the service Node.js is used. Node.js is offering the mechanism for the implementation of frequent I/O bound operations, while also offering integration capabilities for a web client implementation.

While the design of the data analytics workflow is performed in the Analytics and Visualisation Workbench, the execution of the analysis is performed within the Secure and Private Space by interacting with the Resource Orchestrator and the Job Scheduler backend components to ensure the existence of the Secure and Private Space and initiate the application execution (or schedule its execution via the respective interfaces offered by the Job Scheduler). During the execution of the analysis, it interacts via the Master Controller with the SecureSpace Worker and the Decryption Manager before the actual execution in order to ensure the transferring of the selected datasets to the Secure Space. Additionally, the user is able to create an application, which contains the list of datasets that were selected for analysis, the selected algorithm, as well as the selected visualisation type, along with the corresponding parameters, and store it in the platform's application catalogue for later reuse. Furthermore, through the Analytics and Visualisation Workbench and the interactions with the SecureSpace Worker and the Decryption Manager the advanced visualisation capabilities of the platform are offered with a modern data visualisation suite of charts and visualisations that span from basic charts to advanced multilevel visualisations.

**Data Preparation Manager**

The **Data Preparation Manager** is the component that offers the required data manipulation functionalities on top of the available datasets in order to make them suitable for consumption by the Analytics and Visualisation Workbench. The Data Preparation Manager is taking the datasets of the user as input in order to apply a series of data manipulation steps based on the user needs and the results are fed in the Analytics and Visualisation Workbench for data analysis or visualization. The Data Preparation Manager allows the user to create various data preparation jobs by defining on each of them the list of datasets that will be used as the basis for the data preparation and the sequence of data manipulation steps that should be followed in order to transform the original dataset into the one that the user needs.

The main data manipulation functionalities offered can be grouped into: a) column creation (timestamp-related, math-related, aggregation related, shift and conditional operations), b) column drop, c) row filtering, d) column renaming, e) dataset merging, f) null value fill-in operations and g) compute aggregations. The Data Preparation Manager performs all the designed data manipulation process within the Secure Storage as Presto queries and the newly created datasets are also stored with the Secure Storage with the appropriate data access control measures. Hence, the Data Preparation Manager that resides on the Core Platform directly interacts with the Secure Storage towards the preparation of the datasets that will be utilised as input for the data analysis or the visualisation process as designed by the Analytics and Visualisation Workbench.

**SecureSpace Worker**

The **SecureSpace Worker** resides at the Secure and Private Space and is responsible for the job or task execution that is related to the data analysis that is performed on the Secure and Private Space. The SecureSpace Worker is the component that is undertaking the local execution of the jobs as instructed by the Master Controller residing on the Core Platform, utilising the deployed services on the local running environment of the Secure and Private Space. Hence, the SecureSpace Worker is tightly connected with the Master Controller component towards the realisation of the Master / Worker paradigm that is adopted in the ICARUS platform architecture. The SecureSpace Worker is enabling the transfer of the selected encrypted datasets that will be utilised in the data analysis in the Secure Storage. Additionally, it receives a set of instructions that includes the decryption process of the selected datasets as performed by the Decryption Manager, the analytics job execution that is performed by the Jobs Scheduler and Execution Engine and the encryption of the produced results of the analytics as executed by the Encryption Manager. Moreover, the SecureSpace Worker is responsible for providing the encrypted results back to the Secure Storage. For the implementation of the SecureSpace Worker the Spring Roo framework, which provides the easy-to-use rapid development tool for building application in the Java programming language with a large range of features and integration capabilities, is leveraged.

**Jobs Scheduler and Execution Engine**

The **Jobs Scheduler and Execution Engine** is the component in charge of initiating, executing the analytics jobs as designed by the Analytics and Visualisation Workbench, as well as of managing the resources available to the Execution Cluster nodes in the context of a Secure and Private Space. The main functionalities of the Job Scheduler and Execution Engine are: a) the deployment and management of the Execution Cluster by interacting the Resource Orchestrator, b) the execution (immediate or scheduled) of the data analysis on the Execution Engine, and c) the data handling operations related to the access or storage of the data assets that are utilised or produced in the process through the Secure Storage.

Under the hood, the analytics jobs are allocated to the Execution Cluster nodes, decoupling the invocation of a data analysis workflow coming from the Analytics and Visualisation Workbench from its execution. The Jobs Scheduler and Execution Engine is designed as a multi-container service that consists of a job scheduler microservice which is responsible for scheduling the execution of the designed analytics workflows, and the execution engine which parses the definition of an analytic workflow and starts the respective computation. It deploys, scales and manages the nodes involved in analytics jobs execution and interacts with a set of local workers running on the Execution Cluster nodes for distributed computation. The Jobs Scheduler and Execution Engine is also responsible for monitoring the execution of the job and for reporting the execution status to the Analytics and Visualisation Workbench. For the process of loading of datasets or storage of the produced results it is interacting with the Secure Storage, as well as the Encryption Manager for the encryption of the results prior to being stored.

The implementation of the Jobs Scheduler and Execution Engine is based on a customized version of the Spring Cloud Dataflow Server which provides an effective way to execute the designed analytics workflows. Following the micro-service approach, the component allows running algorithms via SpringBoot applications, as well as Apache Spark applications, using an intermediate microservice involved in a pipeline. This intermediate microservice implements a Spark client that interacts with a private Execution Cluster instance that is based on Apache Spark, that is utilised as the cluster computing framework of the platform.

**Execution Cluster**

The **Execution Cluster**, that is managed by the Jobs Scheduler and Execution Engine, is the cluster-computing framework of the platform and is deployed within the Secure and Private Space. The Jobs Scheduler and Execution Engine exploits the capabilities of the Execution Cluster in order to perform the actual analytics workflow execution within the context of a private Execution Cluster instance, guaranteeing the secure and isolated execution.

For the implementation of the Execution Cluster, the Apache Spark has been selected as the cluster-computing framework. Capitalizing on the rich set of features offered by Spark, the Execution Cluster offers the powerful processing engine that enables the data analysis execution across multiple datasets and support the extended list of data analysis algorithms that span from simple statistical analysis to more advance and complex machine learning and deep learning algorithms.

**Encryption Manager**

Within the Secure and Private Space, a running instance of the **Encryption Manager** is deployed. The role of the Encryption Manager in the Secure and Private Space is to encrypt the results of the analysis before they are securely transmitted and stored in the Secure Storage. The Encryption Manager provides the encryption cipher mechanism that generates the symmetric encryption key and the ciphertext that is produced by the encryption of the results. The encryption method that is adopted in the platform is based on a dual encryption approach that follows a symmetric key encryption, based on the AES256 symmetric key encryption algorithm, of the datasets or the results produced by the analysis execution and the secure exchange of a symmetric key between the involved parties upon their agreement for data sharing through secure SSL handshakes during the decryption process. To this end, the Encryption Manager instance of the Secure and Private Space is performing the symmetric encryption of the results interacting with the Job Scheduler and Execution Engine once the data analysis is finished. For the implementation of the Encryption Manager the SpringBoot Java framework is utilised, as well as the Bouncy Castle Java library that is offering a collection of open source lightweight cryptography APIs that complement the default Java Cryptographic Extension (JCE) with extended cryptography functionalities which are suitable for the implementation needs of the Encryption Manager.

**Decryption Manager**

In the same logic as with the Encryption Manager, within the Secure and Private Space a running instance of the **Decryption Manager** is deployed. The role of the Decryption Manager in the Secure and Private Space is to enable the secure and effective decryption of the encrypted datasets on the data consumer side when legitimate access has been obtained without compromising the data privacy of the data provider. Hence, the Decryption Manager is facilitating the reception of the symmetric key that is utilised in the decryption process by interacting with the Encryption Manager through SSL-enabled connection and the decryption of the encrypted dataset with this symmetric key. To this end, the Decryption Manager is involved in the decryption of the selected dataset prior to the execution of the data analytics workflow and the decryption of the results prior to the execution of the visualisation process. The implementation of the Decryption Manager is also based on the SpringBoot Java framework and the Bouncy Castle Java library, similar to the Encryption Manager implementation.

## 3      Results

The ICARUS platform, and the integrated Secure Experimentation Space that were presented in the previous section, have been thoroughly designed and were recently completed in terms of implementation, and thus no scientific results in terms of evaluation of the performance, efficiency, and of course security have yet been made available. Nevertheless, the concept, approach and technical solution will be verified, validated and evaluated through four core representative use cases of the overall aviation's value chain, which are briefly described in the forthcoming sections. The descriptions focus on the scope and expected results of each demonstrator, while a detailed description can be found on the project's deliverable [13].

### 3.1 Extra-aviation services in an integrated airport environment

Airports constitute the most central point in the aviation services environment, as they are the hubs that interconnect passengers, airline companies, tourism organisations, commercial stores as well as city services. The capacity of the airport infrastructure (stands, gates vs planned aircraft arrivals) is generally adequate to meet the demands of the airport users at all times, but during a busy day and especially at peak hours, the demand exceeds the overall capacity of the airport and late arrivals or departures can create significant delays. So, the prime objective of the airport capacity planning is to ensure the most efficient use of the airport infrastructure, to achieve a sustained increase in throughput performance and to increase capacity in all weather conditions.

The aim of the demonstrator is to enable capacity enhancement decisions that are directly targeted to the needs of the airport, but shall also indirectly address interrelated problems that aviation stakeholders operating in the airport (such as airlines and ground handlers) currently face. Through the Secure Experimentation Sandbox of the ICARUS platform, various descriptive and predictive analytics will be conducted to address problems such as: Capacity Modelling, Airport Traffic Forecasting, Flight Delay Prediction, and Position and Slot Allocation / Scheduling, that are all interrelated to the core Airport capacity problem. The expected benefit for the airport is to optimize the Airport airside capacity (including Aircraft parking stands and Passengers Gates) and improve the utilization of all airport airside infrastructure and improve the Runway Operations Capacity. In the early demonstrator activities, the baseline activities for Capacity Modelling and Forecasting (across all business processes: (a) Improved planning of flight schedules per season, (b) Optimum coordination of ground services, (c) Optimization of airport operation services) will be performed. Essentially, they include data assets collection, exploration and experimentation with different analytics algorithms in the early ICARUS platform release. At the same time, the preliminary results for flight delay prediction to contribute to the optimum coordination of ground services will be showcased and evaluated.

### 3.2 Routes analysis for fuel consumption optimization and pollution awareness

Cutting operational expenses while reducing environmental impacts will certainly become among the top challenges of the aviation industry in the next decade. The success of this objective requires improvements not only in the use of resources and materials but also methods and tools. Flexible analysis options support assessing the economic viability of route network extensions or modifications and deliver reliable projections of operational key metrics such as block time, block fuel and payload capacity. In this context two distinct scenarios will be executed within the specific demonstrator: a) Pollution Data Analysis and b) Massive Route Network Analysis and Evaluation utilizing a tool for route analysis, aircraft performance and economic investigations, that will act as the data provider for both scenarios, and the ICARUS platform.

The scope of the first demonstrator scenario comprises a set of activities aiming to support a more accurate analysis of pollution data and aircraft emissions. Typical use cases in this field involve the modelling of pollution data and the prediction of aircraft performance in relation to the environmental impact. The scope of the second demonstrator scenario comprises a set of activities aiming to analyse pollution data on a larger scale, that of a massive route network. Typical use case examples in this field involve the statistical evaluation of weather data, the modelling of aircraft payload capacity scenarios and the prediction of aircraft performance in relation to the underlying route network. In both scenarios, adequate input data shall be compiled in order to conduct the required pre-processing calculations that they will then be processed with suitable ICARUS analytics and linked with other flight information data, if applicable. Finally, the produced results will be visualized with the suitable web dashboards/visualizations that will allow data consumers to review aircraft fuel burn and carbon emissions for defined flight legs or a massive route network. In the early demonstrator activities, the baseline activities for both scenarios for the concept evaluation, specification, prototyping and realisation will be conducted, that includes the data sources collection and pre-processing, the candidate algorithms experimentation and evaluation, and finally the proper visualisation definition.

### 3.3   Aviation-related disease spreading

In the field of computational epidemiology, mathematical models are designed and used together with computational thinking to study the global spreading of epidemics in environments characterized by many degrees of complexity. The modelling tools aim at better understanding various phenomena related to the spread of infectious diseases, to analyse and forecast the evolution of specific epidemic outbreaks to assist policy making in case of public health emergencies. In this context, a meta-population model that uses a data-driven approach based on real-world data on populations and human mobility is currently available. Airline traffic data is a key component for the modelling of human mobility and the simulation of the global spread of an infectious disease and their effect on the population and the economy.

The aim of the demonstrator is to implement non-incremental improvements to the current model by integrating additional available aviation-related data, like travellers' age structure, gender and income data. As a consequence, the specific demonstrator aims at assessing both qualitatively and quantitatively the novel modelling capabilities, analysing the accuracy of the predictions in historical and current epidemic forecasts. In the context of this demonstrator, updated datasets about population and airline traffic, together with official reports about passenger demographics coming from offices of statistics at the country level, will be leveraged to develop an upgraded version of its computational model. At a later stage, the demonstrator shall explore detailed passenger demographics originating from the airline booking systems, and use them to design the modelling framework with a full coupling between human mobility and intra-population interactions. In the early demonstrator activities, the baseline activities for the data exploration and collection, the data pre-processing and cleansing will be performed within the ICAURS platform towards the update of the design of the modelling approach based on the new available data, the simulation code adaptation and data importation in the updated model and finally the updated model validation.

### 3.4 Enhancing passenger experience with ICARUS data

Passenger experience enhancement is a widely discussed topic in aviation, with all involved parties, from airlines and airports to ground handling companies and caterers, looking to optimize their services and product offerings for its achievement. Within the context of this demonstrator, two distinct scenarios will be executed utilizing the capabilities of the ICARUS platform: a) the reduce of cabin food waste towards the increase of revenue and b) the prediction of profitable discounts and offers to increase inflight sales.

The aim of the first scenario of the specific demonstrator is to enhance the existing analytics in the passenger experience by adding prediction capabilities for catering service companies and airlines in order to optimize the loading weight of the duty-free and catering trays on board, prior to the flight, while reducing the cabin food waste. Through this scenario, the implementation of predictive algorithms and methods that will suggest optimized tray loading is expected. The aim of the second scenario of the specific demonstrator is to enhance the existing analytics in the passenger experience by adding prediction capabilities for airlines and catering service companies, in order to suggest discounts and offers targeting to increase in-flight sales. Through this scenario, the targeted predictions and suggestions to airlines and caterers on products and bundles that can be offered on discount towards the increase of in-flight sales and the improvement of the passenger satisfaction and travel experience is expected. In the early demonstrator activities, the baseline activities for both scenarios include the data sources evaluation and availability, the experiment on the analytics algorithms and their evaluation, as well as the verification of the preliminary results from these activities towards their optimisation in the subsequent versions of the demonstrator activities.

## 4 Conclusions

The scope of the current paper is to introduce the ICARUS Secure Experimentation Sandbox of the ICARUS platform. The ICARUS platform aims to provide a big data-enabled platform that aspires to become an "one-stop shop" for aviation data and intelligence marketplace that provides a trusted and secure "sandboxed" analytics workspace, allowing the exploration, curation, integration and deep analysis of original, synthesized and derivative data characterized by different velocity, variety and volume in a trusted and fair manner. To this end, the ICARUS platform offers the ICARUS Secure Experimentation Sandbox as one of its core features that enables the provisioning of a sophisticated environment that can completely guarantees the safety and confidentiality of data, allowing to any interested party to utilise the platform to conduct analytical experiments in closed-lab conditions in order to effectively and securely address the critical barriers for the adoption of Big Data in the aviation industry and enable aviation-related big data scenarios for EU-based companies, organizations and scientists. The concept, approach and technical solution will be verified, validated and evaluated through four core representative use cases of the overall aviation's value chain, as briefly presented.

## 5 Acknowledgement

ICARUS project is being funded by the European Commission under the Horizon 2020 Programme (Grant Agreement No 780792)

## 6 References

1. Wholey, T. J., Deabler, G., & Whitfield, M. M: Commercial Aviation and Aerospace: Big Data Analytics for Advantage, Differentiation and Dollars, Tech. No. GBW03316-USEN-00, Somers, NY: IBM Global Business Services (2014).
2. Cooper, T., Smiley, J., Porter, C., & Precourt, C : Global Fleet & MRO Market Forecast Summary, Oliver Wyman Assessment Report (2016).
3. N. Golchha: Big data-the information revolution, Int. J. Adv. Res. 1(12), 791–794 (2015)
4. C.-W. Tsai : Big data analytics: A survey. *J. Big Data* 2 (1), 1–32 (2015)
5. J.Gantzand, D.Reinsel : The digital universe in 2020 : Big data, bigger digital shadows, and biggest growth in the far east. In: IDC iView: IDC Big Data in 2020, Tech. Rep (2012).
6. Benjelloun, F. Z., & Lahcen, A. A. : Big data security: challenges, recommendations and solutions. In: Web Services: Concepts, Methodologies, Tools, and Applications, pp. 25-38. IGI Global. (2019).
7. Oussous, A., Benjelloun, F. Z., Lahcen, A. A., & Belfkih, S: Big Data technologies: A survey. Journal of King Saud University-Computer and Information Sciences 30(4), 431-448 9 (2018).
8. NIST Big Data Public Working Group: NIST Big Data Interoperability Framework: Volume 4, Security and Privacy Version 2 (No. NIST Special Publication (SP) 1500-4r1), National Institute of Standards and Technology (2018)
9. Venkatraman, S., & Venkatraman, R.: Big data security challenges and strategies. AIMS MATHEMATICS, 4(3), 860-879 (2019).
10. B. Nelson, T. Olovsson: Security and privacy for big data: A systematic literature review. In: 2016 IEEE International Conference on Big Data (Big Data), pp. 3693–3702. (2016)
11. Chidambararajan, B., Kumar, M. S., & Susee, M. S: Big Data Privacy and Security Challenges in Industries. International Research Journal of Engineering and Technology 6 (4), (2019).
12. ICARUS EC H2020 project Homepage, https://www.icarus2020.aero/, last accessed 2020/05/20
13. ICARUS, Demonstrators Execution Scenarios and Readiness Documentation (2019). EC H2020 ICARUS project.